\documentclass[apjl]{emulateapj}

\newcommand{\Tacc}{t_\mathrm{acc}}
\newcommand{\Tsyn}{t_\mathrm{syn}}
\newcommand{\Fen}{F_\mathrm{en}}

\newcommand{\MFA}{magnetic field amplification}
\newcommand{\rgz}{r_{g,0}}
\newcommand{\Beff}{B_\mathrm{eff}}
\newcommand{\kmps}{km s$^{-1}$}
\newcommand{\NL}{nonlinear}
\newcommand{\gamray}{$\gamma$-ray}
\newcommand{\gamrays}{$\gamma$-rays}
\newcommand{\usk}{u_\mathrm{sk}}
\newcommand{\Bsk}{B_\mathrm{sk}}
\newcommand{\Lfeb}{L_\mathrm{FEB}}
\newcommand{\alf}{Alfv\'en}
\newcommand{\SNRJ}{SNR RX J1713.7-3946}
\newcommand{\SC}{self-consistent}
\newcommand{\SCly}{self-consistently}
\newcommand{\Rtot}{R_\mathrm{tot}}
\newcommand{\Rsub}{R_\mathrm{sub}}
\newcommand{\muG}{$\mu$G}
\newcommand{\BISM}{B_\mathrm{ISM}}
\newcommand{\uBwt}{\left <u(x)B(x)\right >}
\newcommand{\Bwt}{\left <B(x)\right >}
\newcommand{\Eemax}{E^\mathrm{max}_e}
\newcommand{\Epmax}{E^\mathrm{max}_p}
\newcommand{\EpmaxUM}{E^\mathrm{max}_p|_\mathrm{UM}}
\newcommand{\EpmaxNL}{E^\mathrm{max}_p|_\mathrm{NL}}

\newcommand{\pNL}{p_\mathrm{NL}^\mathrm{max}}
\newcommand{\pUM}{p_\mathrm{UM}^\mathrm{max}}
\newcommand{\ppmax}{p^\mathrm{max}_p}
\newcommand{\be}{\begin{eqnarray}}
\newcommand{\ee}{\end{eqnarray}}

\newcommand{\rel}{relativistic}

\newcommand{\TP}{test-particle}
\newcommand{\Unmod}{unmodified}

\newcommand{\mc}{Monte Carlo}

\newcommand{\syn}{synchrotron}

\newcommand{\pion}{pion-decay}

\begin{document}

\title{Magnetic Field Amplification and Rapid Time Variations in SNR RX
J1713.7-3946}

\author{Donald C. Ellison\altaffilmark{1} and 
Andrey Vladimirov\altaffilmark{1}} 

\altaffiltext{1}{Physics Department, North
Carolina State University, Box 8202, Raleigh, NC 27695, U.S.A.;
don\_ellison@ncsu.edu, avladim@ncsu.edu}

\begin{abstract}
Evidence is accumulating suggesting that collisionless shocks in
supernova remnants (SNRs) can amplify the interstellar magnetic field to
hundreds of microgauss or even milli-gauss levels, as recently claimed
for \SNRJ. 
If these fields exist, they are almost certainly created by magnetic field
amplification (MFA) associated with the efficient production of cosmic
rays by diffusive shock acceleration (DSA) and their existence
strengthens the case for SNRs being the primary
source of galactic cosmic ray ions to the `knee' and beyond.  However,
the high magnetic field values in SNRs are obtained exclusively from the
interpretation of observations of radiation from relativistic electrons
and if MFA via \NL\ DSA produces these fields 
the magnetic field that determines the maximum ion energy will be
substantially less than the field that determines the maximum electron
energy.
We use results of a steady-state Monte Carlo simulation to
show how nonlinear effects from efficient cosmic ray production and MFA
reduce the maximum energy of protons relative to what would be expected
from test-particle acceleration.
\end{abstract}


\section{Introduction}
Magnetic fields with strengths well above what can be expected from
shock compression of the interstellar medium (ISM) field, $\BISM$, are
inferred in supernova remnants (SNRs) from four types of observations:
(1) sharp X-ray edges or filaments
\citep[e.g.,][]{BambaEtal2003,WarrenEtal2005,CassamEtal2007}, (2)
broadband fits of emission from radio to \gamray\ energies
\citep[e.g.,][]{Cowsik80,VL2003,VBK2005a}, (3) spectral curvature in radio
emission \citep[][]{RE92} and, (4) with the recently reported
{\it Chandra} observations of bright filaments in \SNRJ\
\citep{Uchiyama07}, rapid time variations of nonthermal X-ray emission.
The most likely explanation for these fields is that $\BISM$ is
amplified as part of the efficient production of cosmic rays by
nonlinear diffusive shock acceleration (DSA)
\citep[e.g.,][]{BL2001,AB2006,VEB2006}.  Since the magnetic field
largely determines the maximum particle energy a shock can produce,
\MFA\ (MFA) is critical for interpreting GeV-TeV emission and
determining if SNRs produce cosmic rays to the `knee' and above.
Equally important is the fact that if strong MFA is an intrinsic part of
nonlinear DSA, unexpectedly strong magnetic fields may be present in
other systems undergoing efficient shock acceleration such as radio jets
and gamma-ray burst (GRB) afterglows. The detailed modeling of broadband
observations of SNRs is the only practical way to determine the
importance of \NL\ effects and MFA in high Mach number collisionless
shocks.

Ironically, the evidence for large magnetic fields and, therefore,
nonlinear MFA is obtained exclusively from radiation emitted by \rel\
electrons, while the nonlinear processes responsible for MFA are driven
by the efficient acceleration of \rel\ ions, mainly protons.
Relativistic protons can only be directly inferred from \pion\ emission
at GeV-TeV energies and the unambiguous interpretation of these
observations is difficult and remains uncertain.
In any case, even if \pion\ emission is clearly present, these
observations must be combined with \syn\ emission from electrons to
estimate the magnetic field strength.

A recent report by \citet{Uchiyama07} of rapid time variations ($\sim
1$\,yr) in the nonthermal X-ray emission from bright filaments in \SNRJ\
provides new clues to the electron acceleration process in SNRs.
As emphasized by \citet{Uchiyama07}, the 
rapid time variations
suggest two conclusions.
First, if the variations are determined by the \syn\ radiation losses of
TeV electrons, the magnetic field can indeed be amplified to
milligauss-levels at a SNR shock and, second, the MFA process is tightly
bound with the production of \rel\ particles (i.e., CRs).
An important consequence of these conclusions is that the fraction of
the shock ram kinetic energy that is put into accelerated {\it ions}
must be large, e.g., tens of percent.  Therefore, DSA must be in an
efficient, nonlinear regime where the feedback of accelerated particles
on the shock structure is significant.
A simple estimate confirming this consequence is equation~(15) in
\citet{BL2001}.

Here we address a single question: Can, as asserted by Uchiyama et
al. (2007), the large amplified fields inferred for {\it electrons} from
radiation losses in a \NL\ shock also determine the maximum {\it proton}
energy produced in the SNR shock?  We find the answer to be no because
the inevitable \NL\ shock modification (due to efficient DSA) and the
magnetic field variation in the shock precursor (due to MFA) make the
maximum proton energy smaller than what is expected without accounting
for these effects.  Using results from \citet{VEB2006} for DSA limited
by the finite size of the shock, we estimate that the maximum proton
energy is at least an order of magnitude less than that predicted
assuming an \Unmod\ shock with a large, nearly constant field inferred
from electron radiation losses.  Our result is similar to that found by
\citet{BAC2007} in a time-dependent calculation of DSA where the
acceleration is limited by the age of the shock rather than the size, an
indication that the \NL\ effects we discuss are robust.

\section{Acceleration in a CR Modified Shock}
\label{estimates}
The commonly accepted picture of \NL\ DSA is the following: accelerated
protons diffuse upstream of the shock front, carrying a significant
fraction of the shock ram kinetic energy, and `push' on the incoming
plasma.  In order to conserve momentum, the shock forms a smooth
precursor where the flow speed (in the shock frame) gradually decreases
from the far upstream speed, $u_0$, to a pre-subshock speed, $u_1$,  
and then sharply decreases in a viscous subshock to the downstream
speed, $u_2$. Here, $\Rtot=u_0/u_2$ is the overall compression ratio and
$\Rsub=u_1/u_2$ is the subshock compression ratio.
If MFA accompanies this \NL\ shock smoothing, the low ISM field is
amplified from $B_0=\BISM$ to $B_2$ on the same precursor length
scale.\footnote{Here and elsewhere the subscript ``0'' implies far
  upstream from the shock and ``2'' implies downstream.}

If electrons suffer strong radiation losses, they will have a lower
maximum energy $\Eemax$ than protons and will remain in the vicinity
of the subshock, spending most of their time downstream as they radiate
in the strongly amplified downstream field
$B_2$. Protons, on the contrary, sample the whole precursor and
experience a weaker mean field $\Bwt$, which is
between $\BISM$ and $B_{2}$. Because the escape of protons from the
shock occurs far upstream, where the magnetic field is low, one
can expect that for the most energetic protons, $\Bwt$ is closer to the
weak $\BISM$ than to the strong $B_{2}$, and, consequently, their
maximum energy $\Epmax$ is lower than would be estimated assuming that
the shock is \Unmod\ with $B(x) \sim B_2$ everywhere. The latter case, 
$B(x) \sim B_2$ everywhere, is
implicitly
assumed by \citet{Uchiyama07} when they write 
$\Tacc  = 0.1 (B/\mathrm{mG})^{-1} (E/\mathrm{TeV}) \, \mathrm{yr}$
for the proton acceleration time.

In a size limited shock, the proton maximum energy, $\Epmax$, will be
determined when the upstream diffusion length of the most energetic
protons becomes comparable to the confinement size of the shock,
typically some fraction of the shock radius.  We model the confinement
size with a free escape boundary (FEB) at a distance $\Lfeb$ in front of
the shock. Protons that reach this position stream freely away from the
shock without producing any more magnetic turbulence. Therefore:
\begin{equation}
\label{Ld}
\Lfeb \sim D(\Epmax) / \usk \ ,
\end{equation} 
where for the diffusion coefficient,
$D(E)$, we assume 
\begin{equation}
D(E) = \lambda c/3 = \eta r_g c /3
\ , 
\label{Dp}
\end{equation}
and $\usk$ is the upstream flow speed, $\lambda$ is the scattering
  mean free path, $r_g$ is the gyroradius, and $\eta$ is a parameter
  that characterizes the scattering strength. None of our conclusions
  depend importantly on $\eta$ and we assume `Bohm'
  diffusion and set $\eta=1$. Since for ultra-relativistic protons $r_g
  = E / (eB)$, equations~(\ref{Ld}) and (\ref{Dp}) determine
  $\Epmax$ for a given $\Lfeb$, i.e.,
\begin{equation}
\Epmax \propto \Lfeb  \usk \Bsk 
\ .
\label{Emax}
\end{equation}

For a quasi-parallel, \Unmod\ (UM) shock with no MFA,
the maximum proton energy $\EpmaxUM$ is thus defined by (\ref{Emax})
with $\usk \Bsk = u_0 B_0 = u_0 B_2$, $u_0$ being the shock speed and
$B_2$ being the downstream magnetic field derivable from synchrotron
emission of accelerated electrons.  However, for a \NL\ (NL) CR modified
shock of the same physical confinement size, $\Lfeb$, 
the maximum proton energy $\EpmaxNL$ will be
determined by some mean value $\left<u(x)B(x)\right>$, giving
\begin{equation}
    \frac{\EpmaxNL}{\EpmaxUM}=\frac{\left<u(x) B(x) \right>}{u_0 B_2}
\ .
\end{equation}
For a strongly modified shock, $\left < u(x) B(x)\right > \ll u_0
B_2$, and in the following we determine $\left <
u(x) B(x)\right >/ (u_0 B_2)$ using the Monte Carlo model
described in detail in \citet{VEB2006}.

\section{Monte Carlo Simulation}
The \mc\ model we use \citep[see][ for full details]{VEB2006} calculates
NL DSA and the magnetic turbulence produced in a steady-state,
plane-parallel shock precursor by the CR streaming instability. While
the quasi-linear approximations we make are only valid in the linear
regime we nevertheless follow \citet{BL2001} and assume that the
instability doesn't saturate or damp at $\Delta B \approx B_{0}$ but
continues in the \NL\ regime with $\Delta B \gg B_{0}$.
With this approximation, we are able to \SCly\ determine the \NL\ shock
structure [i.e., $u(x)$ vs. $x$], the MFA [$\Beff(x)$ vs. $x$], and the
thermal particle injection.\footnote{Note that the \mc\ model ignores
the dynamic effects of electrons and the NL shock structure is
determined solely from the backpressure of protons and the magnetic
field. While electron acceleration can be modeled
\citep[e.g.,][]{BaringEtal99}, we only show proton spectra here.}

The NL results we investigate do not depend qualitatively on the
particular shock parameters as long as the sonic Mach number is large
enough to result in efficient DSA.  Here, we use a
shock speed = $u_0 = 3000$\,\kmps,
sonic Mach number $M_s \approx 30$, plasma density 
$n_\mathrm{ISM} = 1$ protons cm$^{-3}$, 
and $B_0=\BISM=10$\,\muG, yielding 
an \alf\ Mach number, $M_A \approx 140$.
To these parameters we add a FEB boundary at $\Lfeb \sim 0.1$\,pc,
corresponding to $10^8\, \rgz$, where $\rgz \equiv m_p u_0 c /(e B_0)$.
This size is comparable to
the hot spots in \SNRJ\ 
and produces a proton energy $\sim 10^{15}$\,eV in our \Unmod\ shock
approximation.

\begin{figure}
\epsscale{.95}
\plotone{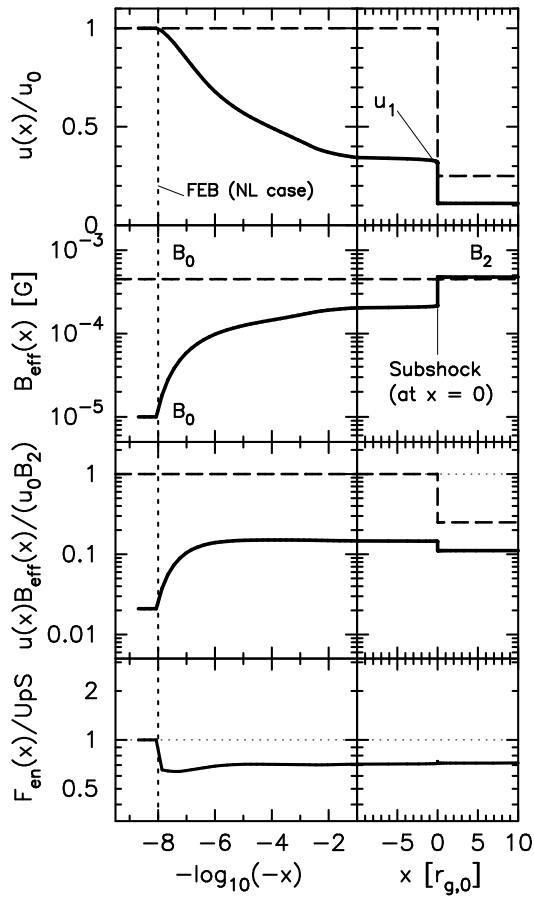}  
\caption{As labeled, the panels show the bulk flow speed in units of the
shock speed, $u_0$, the effective magnetic field, $\Beff$, $[u(x)
\Beff(x)]/(u_0 B_2)$, and the energy flux, $\Fen(x)$, in units of the
far upstream energy flux, all versus position, $x$. The subshock is at
$x=0$ and $\rgz \equiv m_p u_0 c /(e B_0)$ (note the change in $x$-scale
between linear and $\log_{10}$ at $x=-10\rgz$). In all panels, the solid
curves are the \SC, NL result and the dashed curves are for an UM shock
where energy and momentum are not conserved. For the UM shock, the
$x$-independent $\Beff=B_0=B_2 \simeq 450$\,\muG\ has been chosen to
match the amplified field, $B_2$, in the NL result.  Note that the
$x$-scale in physical units depends on $B_0$ which is different in the
two cases but that the FEB is set to the same physical distance in both
cases, i.e., $\Lfeb \sim 0.1$\,pc. The drop in energy flux at the FEB in
the bottom panel results as energetic protons escape the NL shock and
carry away energy. The total energy flux is conserved in the NL case.
\label{shock_profile}}
\end{figure}

\begin{figure}
\epsscale{.95}
\plotone{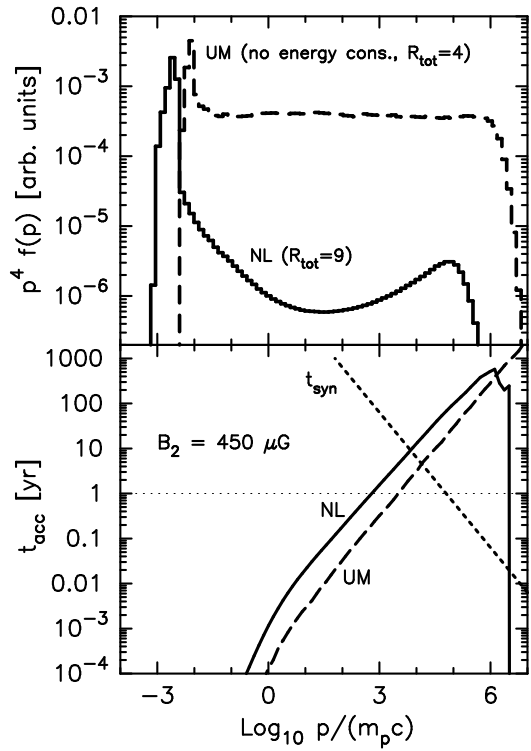}  
\caption{The top panel shows proton spectra, $p^4 f(p)$, and the bottom
  panel shows the acceleration time to a given momentum for the shocks
  shown in Fig.~\ref{shock_profile}. In both panels, the solid curve is
  the NL case and the dashed curve is the UM case.  The maximum momentum
  in the UM case is about a factor of 10 larger than in the NL case. The
  dotted line in the bottom panel is the \syn\ loss time, $\Tsyn \simeq
  12 (B/\mathrm{mG})^{-2} (E/\mathrm{TeV})^{-1}$\,yr, for
  $B=0.45$\,mG. The NL shock accelerates particles at a slower rate than
  the UM shock.
\label{fp}}
\end{figure}

Using the above parameters, we simulate two cases: a \NL\ solution,
where $B$ is amplified from an upstream value $B_0=10$\,\muG\ to a
downstream value $B_2=450$\,\muG\ (obtained self-consistently by our
model), and an \Unmod\ solution with a magnetic field set equal
everywhere to $B_2=450$\,\muG.  In these two cases we look at $\Epmax$
to see how the prediction of the NL model, conserving momentum and
energy, compares to the prediction of the UM model, implicitly assumed
by \citet{Uchiyama07}. The information about the maximum energy of
electrons (which are not included in our calculations) can be inferred
graphically from the plot of the acceleration time (see Fig.~\ref{fp}).

Figure~\ref{shock_profile} shows the shock structure, $u(x)$, the
effective magnetic field after amplification, $\Beff(x)$, and $u(x)
\Beff(x)/(u_0 B_2)$, for the \Unmod\ case (dashed lines), and the \NL\
case (solid lines). The bottom panel shows the energy flux, normalized
to the far upstream value, for the NL case.
The smoothing of $u(x)$, the weak subshock ($\Rsub \simeq
  2.9$), and the increase in $\Rtot$ above 4 ($\Rtot \simeq 9$) are
  clearly present in the top panel for the NL case. These three effects
  must occur to conserve momentum and energy if CRs are efficiently
  accelerated.  The quantity $u(x) \Beff(x)/(u_0 B_2) \sim 0.1$ over
  most of the precursor in the NL case.

In Figure~\ref{fp} we show the momentum phase-space distributions
functions, $f(p)$ (multiplied by $p^4$), and the acceleration time,
$\Tacc$.  The NL effects evident in Fig.~\ref{shock_profile} result in,
\begin{equation}
\pNL / \pUM \lesssim 0.1
\ ,
\end{equation}
and a longer $\Tacc$ to a given momentum. Here, $\pNL = \EpmaxNL/c$ and
$\pUM = \EpmaxUM/c$.
We have not attempted a detailed fit to \SNRJ, but note that
the concave shape of our proton spectrum, above the thermal peak, is
similar to that obtained by \citet{BV2006} who find a good fit to the
data, including the HESS TeV observations
\citep[][]{AharonianJ1713_2006}.

The UM result is obtained from the \mc\ simulation assuming the same
``thermal leakage'' model for injection as in the NL result
\citep[e.g.,][]{JE91}.  This injection scheme works \SCly\ with
modifications in the shock structure and overall compression ratio,
$\Rtot$, to conserve momentum and energy in the NL case. In the UM case,
the shock structure and $\Rtot$ are {\it not} adjusted and the thermal
leakage model produces far too many injected particles to conserve
momentum and energy.
For
the UM shock to become a \TP\ shock with energy conservation, far
fewer particles would need to be injected so that the normalization
of the superthermal $f(p) \propto p^{-4}$ power law becomes low enough,
relative to the thermal peak, so that in contains an insignificant
fraction of the total shock ram kinetic energy.
Since we are only interested in comparing $\ppmax$ in the two cases, the
normalization of the unmodified power law is unimportant since $\pUM$
only depends on $\Lfeb$.

\section{Discussion and Conclusions}
The possibility of strong MFA in SNR shocks has been strengthened by the
recent observations of rapid time variability in hot spots in \SNRJ\ by
\citet{Uchiyama07}.
If we accept the conclusions of \citet{Uchiyama07}, the $\sim 1$~yr
variations in X-ray emission in some hot spots stem from radiation
losses for electrons and indicate magnetic fields on the order of
1\,mG. Such large fields would almost certainly be caused by MFA
occurring simultaneously with the efficient production of CR ions in
DSA.

While a number of other interpretations of the X-ray and broadband
emission in \SNRJ\ have concluded that the magnetic field present in the
particle acceleration site is considerably less than 1\,mG
\citep[e.g.,][]{ESG2001,Lazendic2004,BV2006,PMS2006}, we have shown 
that even if the magnetic field inferred from electron
radiation losses is as high as \citet{Uchiyama07} claim, the underlying
physics of MFA in DSA shows that this field cannot be simply applied to
protons to estimate their maximum energy.

The essential point is that, if MFA to milligauss levels is occurring as
part of DSA, the acceleration must be efficient and the system is
strongly \NL. The accelerated particles and the pressure from the
amplified field feedback on the shock structure
(Fig.~\ref{shock_profile}) and this feedback makes the precursor less
confining [i.e., $\uBwt \ll u_0 B_2$]. Therefore, a shock of a given
physical size will not be able to accelerate protons to an energy as
large as estimated ignoring NL effects.
Even if the proton acceleration is limited by the finite age of the
shock rather that the finite size, as we have assumed, the proton
$\Epmax$ will be less than in a TP approximation because the highest
energy protons will have their acceleration time determined by $\uBwt$
rather than $u_0B_2$. This point has been made by \citet{BAC2007} using
a time-dependent, semi-analytic calculation of DSA with MFA. They have
derived a general equation and determined $\pNL$ for a range of Mach
numbers and choices for the diffusion coefficient when the acceleration
is limited by the shock age.  Considering the differences in the models
and the specific parameters chosen for our examples of young SNRs, the
fact that both calculations yield $\pNL / \pUM \lesssim 0.1$, indicates
the robust nature of the result which should apply in any shock
efficiently producing CRs and magnetic turbulence.

We note that we have neglected the potentially important
physical effect of magnetic field damping in the strong downstream
turbulence \citep[e.g.,][]{Pohl2005}. The importance of damping is still
uncertain \citep[see][]{CassamEtal2007}, but if it occurs it will be
most important for high energy protons with long diffusion lengths. In
effect, damping may determine the FEB but shouldn't qualitatively
change the results for $\pNL/\pUM$.

Despite the reduction in $\Epmax$ compared to \TP\ predictions that
our results imply, a remnant such as \SNRJ\ might still produce CRs up
to the knee.  The NL example we have presented with $B_2 \simeq
450$\,\muG\ produces protons up to $\sim 100$\,TeV in $\sim
100$\,yr in a confinement region of $\sim 0.1$\,pc.  If instead we had
taken $\Lfeb = 1$\,pc, a size comparable to the western shell of \SNRJ,
our NL model would produce $\sim 1$\,PeV protons in $\sim 1000$\,yr.
Protons of this energy are consistent with
the $\sim 30$\,TeV \gamrays\ observed from \SNRJ\
\citep[][]{AharonianJ1713_2006} and when the acceleration of heavy ions
such as Fe$^{+26}$ is considered, the maximum particle energy extends to
$> 10^{16}$\,eV.

The production of turbulence in efficient DSA is an active area of
research and the question arises if non-resonant wave generation will
substantially modify our results. For instance, recently
\citet{PLM2006} have investigated Bell's non-resonant turbulence
generation mechanism \citep{Bell2004} and found that it can dominate
over resonant instabilities for high speed shocks (e.g., $\usk \simeq
0.1 c$). They also find, however, that for slower shocks of the order we
assume here ($\usk \sim
0.01 c$), resonant instabilities dominate. The \citet{VEB2006} model
assumes resonant instabilities.

Other effects may be important as well such as particles escaping at the
FEB.  In a size limited shock, the escaping energy flux can be large if
DSA is efficient. For our example, the bottom panel of Fig.~\ref{fp}
shows that $\sim 30$\% of the energy flux escapes at the upstream FEB
and this highly anisotropic beam of particles with energies near
$\Epmax$ moving through the cold ISM should efficiently generate
turbulence. 
The details of this so-called ``magnetic bootstrap'' scenario
\citep[e.g.,][]{BF2007} are just starting to be investigated but such
turbulence might generate higher effective fields with a correspondingly
higher $\Epmax$.
Despite these possibilites and the uncertainties involved in wave
generation, we believe none of these effects will qualitatively change
our conclusions: the precursor field must still range from $B_0$ to
$B_2$ and the proton maximum energy will be determined by a smaller
effective field than that estimated from \syn\ losses for electrons.

As a final comment we emphasize a point also made by \citet{BAC2007}. If
MFA is occurring and the system is highly NL, it may not be possible to
explain temporal variations in nonthermal X-ray emission
simply as a radiation loss time.
There cannot be variations in X-ray emission on
short time scales unless the accelerator changes in some fashion on
these time scales, otherwise the radiation would be steady, or varying
on the shock dynamic timescale, regardless of
how short the radiation loss time was.
Since the injection and acceleration of protons and electrons is
nonlinearly connected to the amplified magnetic field,
changes in the
electron particle distribution and changes in the field producing
the \syn\ emission, will go together and it may be difficult to
unambiguously determine the field strength from temporal variations.

\acknowledgments 
We thank R. Blandford, Y. Butt, and S. Funk for helpful
discussions and acknowledge support from NASA grants ATP02-0042-0006,
NNH04Zss001N-LTSA, and 06-ATP06-21. D.C.E. wishes to thank the Kavli
Institute for Particle Astrophysics and Cosmology (KIPAC) where part of
this work was done.

\end{document}